# IDR Muon Capture Front End and Variations

David Neuffer[a], Gersende Prior[b], Christopher Rogers[c], Pavel Snopok[d] and Cary Yoshikawa[e]

[a]Fermilab, PO Box 500, Batavia IL 60510
[b]CERN, Geneva, Switzerland
[c]ASTeC, Rutherford Appleton Laboratory OX11 0QX, UK
[d]Physics Dept., University of California, Riverside, CA
[e]Muons, Inc., Batavia IL 60510

**Abstract.** The (International Design Report) IDR neutrino factory scenario for capture, bunching, phase-energy rotation and initial cooling of μ's produced from a proton source target is explored. It requires a drift section from the target, a bunching section and a φ-δE rotation section leading into the cooling channel. The rf frequency changes along the bunching and rotation transport in order to form the μ's into a train of equal-energy bunches suitable for cooling and acceleration. Optimization and variations are discussed. An important concern is rf limitations within the focusing magnetic fields; mitigation procedures are described. The method can be extended to provide muons for a $\mu^+$-$\mu^-$ Collider; variations toward optimizing that extension are discussed.

**PACS:** 29.25.-t, 29.27.-a, 14.60.Pq. http://www.aip.org/pacs/index.html

## INTRODUCTION

In a neutrino factory, short, intense bunches of protons are focused onto a target to produce pions, which decay into muons and are then accelerated into a high-energy storage ring, where their decays provide beams of high-energy neutrinos.[1, 2, 3] The challenge is to collect and accelerate as many muons as possible. The IDR neutrino factory consists of:

- a proton source with an intensity of 4MW beam power (50Hz, 8 GeV protons, ~2ns bunches). (~6×10$^{13}$p/ cycle, up to ~3 bunches/cycle),
- a target, capture and cooling section that produces π's that decay into μ's and captures them into a small number of bunches,
- an accelerator that takes the μ's to 25 GeV for insertion into storage rings. μ decay in the straight sections of storage rings provides high-energy ν beams for:
- ~100 kton ν-detectors placed at 4000-7500km baselines with the ability to identify ν-interactions.

The goal is > 10$^{21}$ν's/beamline/year in order to obtain precise measurements of ν-oscillation parameters.

## FRONT END OVERVIEW

This paper describes the Front End of the neutrino factory system, which consists of the target and capture and cooling section that takes the π's produced at the target, and captures and cools the resulting decay μ's, preparing them for insertion into the μ accelerators. Fig. 1 shows an overview of the system. The present system is a modification of similar systems included in the neutrino factory study 2a and the ISS study.[4]

### Target And Decay Channel

The 5-15 GeV proton source produces short pulses of protons that are focused onto a liquid-Hg-jet target immersed in a high-field solenoid with an internal beam pipe radius $r_{sol}$. The proton bunch length is 1 to 3 ns rms (~5 to 15 ns full-width), $B_{sol}$=20T, and $r_{sol}$ = 0.075m. Secondary particles are radially captured if they have a transverse momentum $p_T$ less than ~$ecB_{sol}r_{sol}/2$ = 0.225 GeV/c. Downstream of the target solenoid the magnetic field is adiabatically reduced from 20T to 1.5T over a distance of ~15m, while the beam pipe radius increases to 0.3m. This arrangement captures within the B=1.5T decay channel a secondary pion beam with a broad energy spread (~50 MeV to 400 MeV kinetic energy).

The initial proton bunch is relatively short, and as the secondary pions drift from the target they spread apart longitudinally, following: $c\tau(s) = s/\beta_z + c\tau_0$, where s is distance along the transport and $\beta_z = v_z/c$ is the longitudinal velocity. Hence, downstream of the target, the pions and their daughter muons develop a position-energy correlation in the rf-free decay channel. In the IDR baseline, the drift length $L_D$ = 64.6m, and at the end of the decay channel there are about 0.2 muons of each sign per incident 8 GeV proton.

## rf Buncher

The drift channel is followed by a buncher section that uses rf cavities to form the muon beam into a train of bunches, and a phase-energy rotating section that decelerates the leading high energy bunches and accelerates the late low energy bunches, so that each bunch has the same mean energy. The IDR design delivers a bunch train that is <80m long, which is an improvement over the version of the design developed for the ISS [16, 12] which delivered an 120m long bunch train containing the same number of muons.

To determine the required buncher parameters, we consider reference particles (0, N) at $P_0$= 233 MeV/c and $P_N$ = 154 MeV/c, with the intent of capturing muons from a large initial energy range (~50 to ~400 MeV). The rf frequency $f_{rf}$ and phase are set to place these particles at the center of bunches while the rf voltage increases along the transport. These conditions can be maintained if the rf wavelength $\lambda_{rf}$ increases along the buncher, following:

$$N_B \lambda_{rf}(s) = N_B \frac{c}{f_{rf}(s)} = s\left(\frac{1}{\beta_N} - \frac{1}{\beta_0}\right)$$

where s is the total distance from the target, $\beta_1$ and $\beta_2$ are the velocities of the reference particles, and N is an integer. For the present design, N is chosen to be 10, and the buncher length is 33m. With these parameters, the rf cavities decrease in frequency from ~320 MHz ($\lambda_{rf}$ = 0.94m) to ~230 MHz ($\lambda_{rf}$ = 1.3m) over the buncher length.

The initial geometry for rf cavity placement uses 0.5m long cavities placed within 0.75m long cells. The 1.5T solenoid field focusing of the decay region is continued through the buncher and the following rotator section. The rf gradient is increased from cell to cell along the buncher, and the beam is captured into a string of bunches, each of them centered about a test particle position with energies determined by the $\delta(1/\beta)$ spacing from the initial test particle:

$$1/\beta_n = 1/\beta_0 + n\ \delta(1/\beta),$$

where $\delta(1/\beta)=(1/\beta_N - 1/\beta_0)/N$. In the initial design, the cavity gradients follow a linear increase along the buncher:

$$V'_{rf}(z) \approx 9\left(\frac{z}{L_{Bf}}\right) MV/m$$

where z is distance along the buncher. The gradient at the end of the buncher is 9 MV/m. This gradual increase of the bunching voltage enables a somewhat adiabatic capture of the muons into separated bunches, which minimizes phase space dilution.

In the practical implementation of the buncher concept, this linear ramp of changing frequency cavities is approximated by a sequence of rf cavities that decrease in frequency along the 33m beam transport allotted to the buncher. A total of 37 rf cavities are specified, with frequencies varying from 319.6 to 233.6, and rf gradients from 4 to 7.5 MV/m The number of different rf frequencies is limited to a more manageable 13 (~3 rf cavities per frequency). The linear ramp in gradient is approximated by the placement and gradient of the cavities in the buncher. Table 1 lists the rf cavity requirements. At the end of the buncher, the beam is formed into a train of positive and negative bunches with different energies.

## Phase-Energy Rotator

In the rotator section, the rf bunch spacing between the reference particles is shifted away from the integer $N_B$ by an increment $\delta N_B$, and phased so that the high-energy reference particle is stationary and the low-energy one is uniformly accelerated to arrive at the same energy as the first reference particle at the end of the rotator. For the IDS $\delta N_B$ =0.05 and the bunch spacing between the reference particles is $N_B + \delta N_B$ = 10.05. This is accomplished using an rf gradient of 12 MV/m in 0.5m long rf cavities within 0.75m long cells. The rf frequency decreases from 230.2 MHz to 202.3 MHz from cavity to cavity down the length of the 42m long rotator region. As in the buncher, the number of independent rf frequencies is held to a more manageable number by running several cavities at the same frequencies. Table 1 summarizes the rf cavity requirements. The rotator uses 54 rf cavites of 15 different frequencies.

Within the rotator, as the reference particles are accelerated to the central energy (at P=233MeV/c) at the end of the channel, the beam bunches formed before and after the central bunch are decelerated and accelerated, respectively, obtaining at the end of the rotator a string of bunches of approximately equal energies. (Both positive and negative muons form a string of equal-energy bunches.) The particle bunches are then aligned with nearly equal central energies. At the end of the rotator the rf frequency matches into the rf frequency of the ionization cooling channel (201.25 MHz). The average momentum at this exit is ~250 MeV/c. The performance of the bunching and phase rotation channel, along with the subsequent cooling channel, is displayed in Fig. 3, which shows, as a function of the distance down the channel, the number of muons within a reference acceptance (muons within 201.25 MHz rf bunches with transverse amplitudes less than 0.03m and longitudinal amplitudes less than 0.15m). The phase rotation increases the "accepted" muons by a factor of 4.

A critical feature of the muon production, collection, bunching and phase rotation system that we have described is that it produces bunches of both signs ($\mu^+$ and $\mu^-$) at roughly equal intensities. This occurs because the focusing systems are solenoids which focus both signs, and the rf systems have stable acceleration for both signs, separated by a phase difference of $\pi$.

## Cooling Channel

The IDS baseline cooling channel design consists of a sequence of identical 1.5m long cells (Fig. 2). Each cell contains two 0.5m-long rf cavities, with 1.1cm thick LiH blocks at the ends of each cavity (4 per cell) and a 0.25m spacing between cavities. The LiH blocks provide the energy loss material for ionization cooling. The cells contain two solenoidal coils with the coils containing opposite sign currents. The coils produce an approximately sinusoidal variation of the magnetic field in the channel with a peak value on-axis of ~2.8T, providing transverse focusing with $\beta_\perp \cong 0.8$m. The currents in the first two cells are perturbed from the reference values to provide matching from the constant-field solenoid in the buncher and rotator sections. The total length of the cooling section is ~75m (50 cells). Based on simulations, the cooling channel is expected to reduce the rms transverse normalized emittance from $\varepsilon_{N,rms}$ = 0.018m to $\varepsilon_{N,rms}$ = 0.007m. The rms longitudinal emittance is $\varepsilon_{L,rms}$ = ~0.07 m/bunch.

The $\mu$ accelerator will have a restricted aperture. From simulation studies, the acceptance is approximated by the estimate that $\mu$'s with transverse amplitudes < 0.03m and longitudinal amplitudes < 0.15m (as defined in ref. 5) will be accelerated in the $\nu$-factory. This corresponds to a restricted subset of the $\mu$'s produced and transported in the Front End system. In simulations this accepted beam has a smaller emittance. The rms transverse emittance is $\varepsilon_{N,rms} \cong 0.004$m, and the longitudinal emittance is ~0.036m/bunch.

At the end of the cooling channel there are interlaced trains of bunches of $\mu^+$ and $\mu^-$, of similar intensity. The trains of useable muons are ~80m long (~54 bunches). Each of the bunches has an rms length of 0.15m, and an rms momentum width of ~30MeV/c, with a mean momentum of ~250MeV/c. ~0.1 $\mu^+$ and $\mu^-$ per initial 8 GeV proton are within acceptances.

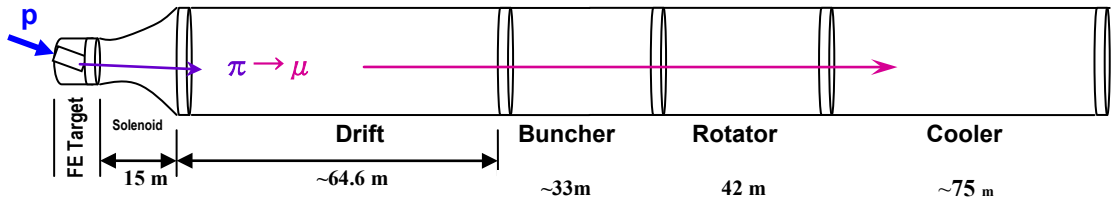

**FIGURE 1.** : Overview of the IDS neutrino factory front end, consisting of a target solenoid (20 T), a tapered capture solenoid (20 T to 1.5T, 15 m long), Drift section (64.6 m), rf Buncher (33 m), an energy-phase Rotator (42 m), and a Cooler (75 m).

**TABLE 1.** Summary of front end rf requirements

| Region | Length (m.) | Number of Cavities | Number of frequencies | Frequencies [MHz] | Peak Gradient {MV/m} | Peak Power requirements |
|---|---|---|---|---|---|---|
| Buncher | 33 | 37 | 13 | 319.6 to 233.6 | 4 to 7.5 | 1 to 3.5 MW/freq. |
| Rotator | 42 | 56 | 15 | 230.2 to 202.3 | 12 | 2.5 MW/cavity |
| Cooler | 75 | 100 | 1 | 201.25 | 15 | 4 MW/cavity |
| Total (with Drift) | 230m | 193 | 29 | 319.6 to 201.25 | 1000MV | 550MW |

## SIMULATION STUDIES AND VARIATIONS

The front end described here has been simulated using the simulation codes ICOOL[6] and G4BeamLine.[7] Both of these are particle tracking codes that use semi-analytic methods to calculate magnetic and rf fields. They include particle decay, ionization energy loss, multiple scattering and energy straggling, and they attempt to calculate the particle motion and interactions in realistic and accurate models.

Fig. 3 shows results of simulations, using both ICOOL and G4Beamline. In these simulations, an initial set of secondary particles is generated at the production target, using a model of 8 GeV proton beam on a Hg jet target, using MARS 15.07.[8] The MARS-generated secondaries are tracked through the drift buncher and phase rotator, and cooler. π's decay to μ's, and are bunched and cooled.

The two simulations are in close agreement, an important verification of the simulation codes. The actual uncertainty is probably greater; different versions of the same code vary by ~10%.

## rf Gradients in Magnetic Fields

The present configuration requires relatively high gradient rf cavities within strong solenoidal fields (~12MV/m at ~200 MHz within ~1.5T in the Rotator, and ~15MV/m within alternating solenoid fields in the Cooler). Experimental studies have shown that rf cavities within magnetic fields breakdown at smaller gradients than rf within field-free regions. For example, a 805 MHz pillbox rf cavity that operated at 40MV/m in field-free vacuum was reduced to ~15 MV/m within a 3T solenoid.[9] Experimental study at the Front End parameters has not yet been attempted, but rf gradients may be limited to less than the current design values. rf studies at those parameters will be attempted soon.

The Front End design can be adapted to rf gradient limitations as they are discovered. A simplest adaptation is to maintain the initial design but use rf at reduced gradient values. Studies of the effects of such reduced gradient operation have been made. Reduction of baseline gradients by as much as a factor of 2 would only reduce μ acceptance by ~30%. Reduction by more than a factor of 2 would probably require more fundamental changes in the Front End design. Alternative designs that directly reduce the magnetic field and rf effects have also been developed and are described next.

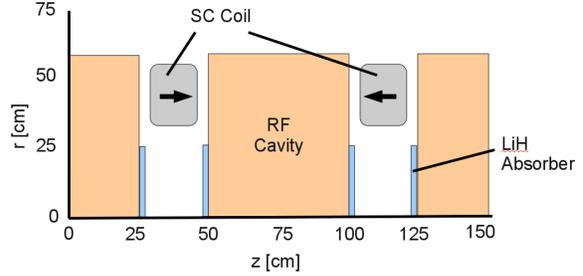

**Figure 2**: r-z overview of a 1.5m long cooling cell showing rf cavities, LiH absorbers and magnetic coils. The structure is cylindrically symmetric.

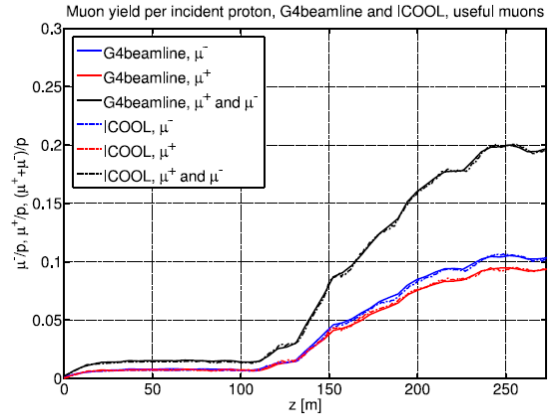

**Figure 3**: Results of simulations. This figure displays the evolution of the number of $\mu^+$ and $\mu^-$ within the reference acceptances ($A_t < 0.03$m, $A_L < 0.15$m) That number increases from ~0.04 $\mu^+$/p to ~0.1 $\mu^+$/p over the cooling channel (~160m to ~240m), with a similar number of $\mu^-$.

## Alternative Front End Lattices

Experiments have shown that high-density $H_2$ gas suppresses breakdown in rf cavities and that suppression occurs even within strong magnetic fields. Addition of gas is particularly desirable in the cooling section where the rf gradients are greatest and also where the gas is also a beam-cooling energy absorber. It has been noted that introduction of $H_2$ gas at the level of ~10 atmospheres (at room temperature) into the cooling lattice of Fig. 2 should be sufficient to suppress breakdown and simulations show that μ cooling and acceptance would be the same as the baseline. This relatively manageable pressure could be introduced into the Front End with minimal design changes.[10]

In this initial redesign the Buncher and Rotator cavities were unchanged, although they operate at lower gradient than the Cooler. Gas for breakdown suppression could be introduced in these as well. rf gradients would need to be increased and phases rematched to obtain similar bunching/rotation results.

If the $H_2$ gas density were increased to ~70 atmospheres, the gas could completely replace the LiH absorbers. Beam cooling would actually be improved, because $H_2$ has less multiple scattering than LiH. This would require a more difficult, and perhaps expensive, high-pressure redesign of the front end, however. There is a concern that ionization electrons from the intense muon beams in the front end would drain energy in gas-filled cavities; an rf experiment has been initiated that will test this potential difficulty.

The magnet lattices can be redesigned such that the magnetic fields at the rf cavities are very small (< ~0.1T). In a magnetically shielded lattice the magnetic field at the rf cavities is reduced by stretching the lattice and adding iron shielding around the coil. (see Fig. 4) This reduces the acceptance of the transport and requires higher rf gradients to obtain useful cooling and acceptance.[11] Studies on this reconfiguration are continuing.

It is possible that breakdown is more likely if magnetic fields are perpendicular to the rf surface and less likely when they are parallel to the surface. The rf cavities and the magnetic coil placement could be redesigned so that the rf cavity surfaces are parallel to the magnetic fields. (see Fig. 5) Stratakis et al.[12] have designed and simulated a Front End with a cooling lattice with these properties (called "magnetic insulation" of rf), and obtain similar acceptance to the baseline. The resultant rf cavity shapes require twice as much rf power as the pill box rf cavities.

## Beam Losses and Shielding

In the front end, ~4MW of proton beam is placed on target, producing ~25kW of ~200MeV μ's that are captured, along with a large number of uncaptured secondaries. This is likely to deposit a large amount of secondary particle energy throughout the Front End system. In currently operating accelerators, uncontrolled hadronic losses must be less than ~1W/m to enable "hands-on" maintenance without additional time/distance/shielding constraints. Beam losses must also be separated from superconducting magnet coils. Fig. 6 shows secondary particle losses through the channel in ICOOL simulations, where the losses are generated by the program aperture constraints, without any shielding or collimation for loss control. In much of the transport, losses are at the ~100W/m scale and demonstrate a need for more careful consideration of particle loss effects.

More shielding in the front end can be inserted; in the first 70 m we have a ~30cm beam pipe within ~65 cm. solenoid coils. Shielding between these could control losses in that region. Hadronic absorbers (with momentum selecting chicanes) can also be inserted to control loss. The cooling energy absorbers are both a source and an absorber of secondaries. An integrated evaluation of the front end, including beam loss effects and mitigation consequences, is needed.

## ADAPTATION TO MUON COLLIDER FRONT END

In current planning for a muon collider, the initial μ production and capture design would be the same as for the IDS ν-factory. The collider also needs to collect a maximal number of μ's, and the ability to capture both $\mu^+$ and $\mu^-$ simultaneously is required. A key difference is that the muon bunches must be recombined into single μ+ and μ- bunches; also, the beam must be be cooled both transversely and longitudinally to much smaller emittances.

In reoptimizing for a collider, we capture in a shorter bunch train by increasing the capture energy, increasing rf gradients, and using a shorter front end. We assume that a second-generation Front End could use higher gradients, and longitudinal cooling increases longitudinal acceptance. In an incremental redesign, we chose to collect at 275MeV/c, increased rf gradients from 9/12/15 MV/m to 15/16/18 MV/m and reduced the Front End length by ~30m. With this redesign, we obtain ~0.12 $\mu^+$ and $\mu^-$ /p in trains of ~12 bunches, suitable for downstream cooling and recombination. Bunch recombination can be obtained by a (~ time-reversed) adaptation of the present front end buncher approach, combined with longitudinal cooling. A helical transport with enhanced chronicity is helpful in recombination.[13]

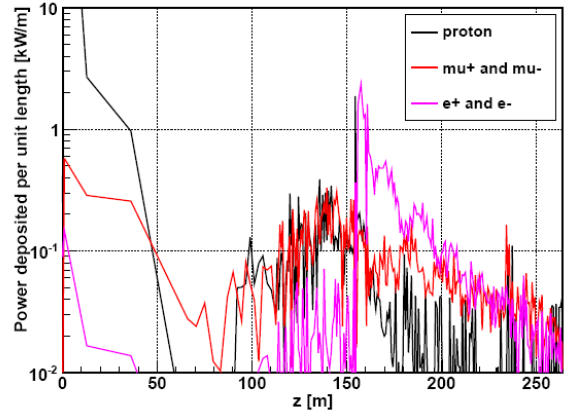

**Figure 6**: Power deposited in the IDS front end due to particle loss (traversing apertures), obtained using the ICOOL code.

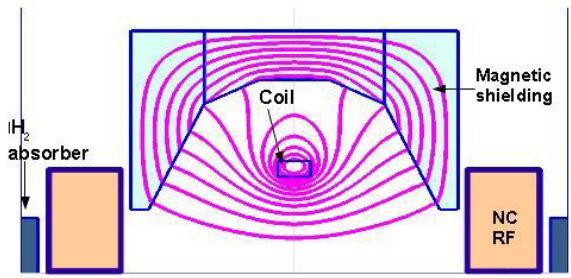

**Figure 4.** Overview (r-z) of a magnetically-shielded rf lattice cell, showing focusing coil, shielding, rf and absorbers. The space required for shielding limits the space for rf and higher gradient is needed to compensate. Cell is cylindrically symmetric.

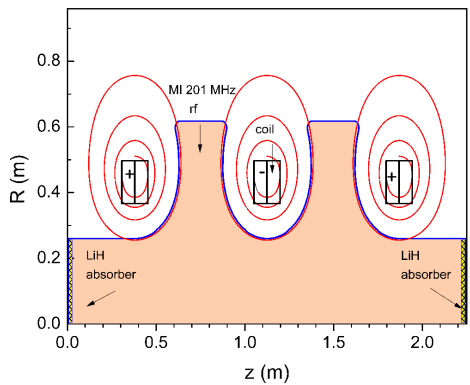

**Figure 5** :r-z cylindrical cross section of a cooling cell with magnetically insulated rf cavities.